\documentclass[12pt]{article}
\usepackage{amsmath,amssymb}
\newcommand{\be}{\begin{equation}}
\newcommand{\ee}{\end{equation}}
\newcommand{\bea}{\begin{eqnarray}}
\newcommand{\eea}{\end{eqnarray}}

\def \th {$\theta_{\mu\nu}$}
\def\P{Poincar\'e }

\begin{document}
\renewcommand {\theequation}{\thesection.\arabic{equation}}
\renewcommand {\thefootnote}{\fnsymbol{footnote}}
\vskip1cm
\begin{flushright}
\end{flushright}
\vskip1cm
\begin{center}
{\Large\bf Twisted Poincar\'e Symmetry and Some
Implications\\\vskip0.3cm on Noncommutative Quantum Field Theory}
\footnote{Talk given at the 21st Nishinomiya-Yukawa Memorial
Symposium on Theoretical Physics "Noncommutative Geometry and
Quantum Space-Time in Physics", Nishinomiya-Kyoto, Japan, 11-15
November 2006; to appear in Prog. Theor. Phys. Suppl.}

\vskip 1cm {\large\bf Anca Tureanu}

\vskip .7cm

{\it $^{a}$High Energy Physics Division, Department of Physical
Sciences,
University of Helsinki\\
\ \ {and}\\
\ \ Helsinki Institute of Physics,\\ P.O. Box 64, FIN-00014
Helsinki, Finland}
\end{center}
\vskip1cm
\begin{abstract}

The concept of twisted Poincar\'e symmetry, as well as some
implications, are reviewed. The spin-statistics relation and the
nonlocality of NC QFT are discussed in the light of this quantum
symmetry. The possibility of a twisted symmetry principle for
quantum field and gauge theories formulated on a noncommutative
space-time is also explored.

\end{abstract}

\vskip1cm


\section{Twisted Poincar\'e symmetry}\label{twist-P}

The study of the quantum space-time, described by the
Heisenberg-like commutation relation
\be\label{cr}[\hat x_\mu,\hat x_\nu]=i\theta_{\mu\nu}\,,\ee
where $\theta_{\mu\nu}$ is an antisymmetric matrix, is strongly
motivated by string theory \cite{SW}, as well as the merging of
Heisenberg's principle with Einstein's theory of gravity \cite{DFR}.
The traditional framework for constructing quantum field theory
(QFT) on such a noncommutative (NC) space-time has been the
Weyl-Moyal correspondence, in which to each field operator
$\Phi(\hat x)$ corresponds a Weyl symbol $\Phi(x)$, defined on the
commutative counterpart of the space-time. Also, in the action
functional, the products of field operators, e.g. $\Phi(\hat
x)\Psi(\hat x)$ is replaced by the Moyal $\star$-product of Weyl
symbols, $\Phi(x)\star\Psi(x)$, where
\be\label{star}\star=\exp{\left(\frac{i}{2}\theta_{\mu\nu}\overleftarrow\partial^\mu\overrightarrow\partial^\nu\right)}\
.\ee
In this correspondence, the operator commutation relation (\ref{cr})
becomes
\be\label{star cr}[ x_\mu, x_\nu]_\star=x_\mu\star x_\nu-x_\nu\star
x_\mu=i\theta_{\mu\nu}\ee
and NC models have been built by taking their commutative
counterparts and replacing the usual multiplication by
$\star$-product.

Such NC field theories lack Lorentz invariance, while the
translational symmetry is preserved. On the 4-dimensional space-time
there exists a frame in which the antisymmetric matrix
$\theta_{\mu\nu}$ takes the block-diagonal form:
\begin{eqnarray}
\theta^{\mu\nu}=\left(
\begin{array}{cccc}
0 & \theta & 0  & 0 \\
-\theta & 0 & 0  & 0 \\
0 & 0  &0 & \theta' \\
0 & 0  & -\theta' & 0
\end{array}
\right). \nonumber
\end{eqnarray}
 The Lorentz group
is thus broken to its $SO(1,1)\times SO(2)$ subgroup (or to
$O(1,1)\times SO(2)$, if according to the present-day wisdom, one
considers $\theta_{0i}=0$), and one consequently encounters a
problem with the representations: both $O(1,1)$ and $SO(2)$ being
Abelian groups, they have only one-dimensional unitary irreducible
representations and thus no spinor, vector etc. representations.

The problem with the representations had been a standing one until
it was realized that NC QFTs possess, however, a quantum symmetry
(see the monographs \cite{ChPr,Maj,ChD}), namely under the {\it
twisted \P algebra} \cite{CKT} (see also Ref. \cite{oeckl} for a
dual approach), which is the universal enveloping of the \P algebra
$\cal U(\cal P)$ viewed as a Hopf algebra deformed with the Abelian
twist element \cite{drinfeld}
\be\label{abelian twist}{\cal
F}=\exp\left({\frac{i}{2}\theta^{\mu\nu}P_\mu\otimes
P_\nu}\right).\ee
This induces on the algebra of representation of the \P algebra the
deformed multiplication,
\be\label{twist prod} m\circ(\phi\otimes\psi)=\phi\psi\rightarrow
m_\star\circ(\phi\otimes\psi)=m\circ{\cal
F}^{-1}(\phi\otimes\psi)\equiv \phi\star\psi\,,\ee
which is nothing else but the $\star$-product (\ref{star}).

The twist (\ref{abelian twist}) does not affect the actual
commutation relations of the generators of the \P algebra $\cal P$.
However it deforms the action of the generators in the tensor
product of representations, or the so-called {\it coproduct}. In the
case of the usual \P algebra, the coproduct $\Delta_0\in\cal U(\cal
P)\times \cal U(\cal P)$ is symmetric,
$$\Delta_0(Y)=Y\otimes1+1\otimes Y,$$ for all generators $Y\in \cal
P$. The twist $\cal F$ deformes the coproduct:
\be\label{twist_coproduct}\Delta_0(Y)\longmapsto\Delta_t(Y)={\cal
F}\Delta_0(Y){\cal F}^{-1}\,.\ee
This similarity transformation is compatible with all the properties
of $\cal U(\cal P)$ as a Hopf algebra, since $\cal F$ satisfies the
twist equation:
\be\label{twist_eq} {\cal F}_{12}(\Delta_0\otimes id){\cal F}={\cal
F}_{23}(id\otimes\Delta_0){\cal F}\,. \ee

The deformation of the coproduct requires, for consistency, a
corresponding deformation of the multiplication (\ref{twist prod})
in the algebra of representations. Taking in (\ref{twist prod})
$\phi(x)=x_\mu$ and $\psi(x)=x_\nu$, one obtains:
\be [x_\mu, x_\nu]_\star=i\theta_{\mu\nu}. \ee
Thus, the construction of a NC quantum field theory through the
Weyl-Moyal correspondence is equivalent to the procedure of
redefining the multiplication of functions, so that it is consistent
with the twisted coproduct of the \P generators
(\ref{twist_coproduct}). The emerging NC QFT is invariant under the
twisted \P algebra.

The main physical implication is that the representation content of
the twisted \P algebra is identical to the usual \P algebra, because
the commutation relations of the generators themselves are exactly
the same. This justifies all the results which had been previously
obtained in NC QFT, using the representations of the usual \P
algebra. Moreover, the twisted \P symmetry implies a new concept of
relativistic invariance for the NC space-time \cite{CPrT}.

The formulation of this particular form of noncommutativity of
space-time as a quantum symmetry was first done in the context of
noncommutative string theory and quasitriagular Hopf algebras
\cite{Watts} and soon afterwards in the dual language to Hopf
algebras \cite{oeckl}, but overlooking the transparent approach of
twisting the universal enveloping algebra of the \P algebra, as well
as the implications on the field representations.

The twisted Poincar\'e symmetry sheds a new light on the issues in
which the representations of the fields are important. It is of
course interesting to see whether it also leads to some
fundamentally new results. Many interesting developments have been
already investigated: the twisted \P symmetry has been extended to
the supersymmetric case \cite{KS}, the relation between the global
$\theta$-\P deformation and the explicit symmetry breaking has been
established \cite{Gonera}, but probably the most studied aspect at
the moment is the noncommutative gravity formulated by twisting the
algebra of diffeomorphisms \cite{Wess-grav}.\footnote{It is however
puzzling that the general coordinate transformations are deformed
with the frame-dependent twist (\ref{abelian twist}), leading to a
usual $\star$-product, which does not transform under general
coordinate transformations. This approach does not match the
string-theoretical one \cite{AMV}, therefore the problem deserves a
further, perhaps more sophisticated, treatement.}

In the following we shall consider two crucial aspects of NC QFT:
the spin-statistics relation - which is the basis for the stability
of matter in the Univers, and whose possible violation would have
major implications, and a new possibility of formulating
noncommutative gauge theories - an essential ingredient for any
plausible model building. The latter aspect will be considered in a
more general framework of a possible twisted symmetry principle for
NC QFT and gauge theories.

\section{The spin-statistics relation in NC QFT}\label{spin statistics}

The spin-statistics relation in NC QFT was proven to be valid both
using Pauli's approach \cite{CNT} and in the axiomatic formulation
based on the $O(1,1)\times SO(2)$ symmetry group \cite{axiomatic}.

The quantum group approach to NC QFT fully supports these previous
results. To show this, we shall make a short digression to braided
groups. Quantum groups (of which the twisted \P algebra is an
example) are not correlated to any kind of statistics, in the sense
that the permutation of different copies of the same objects do
satisfy usual bosonic/fermionic commutation rules. Another possible
line of development is to use the idea of the deformation
(generalization) of the very permutation rule ("statistics"),
leading to the so-called braided groups (see Ref. \cite{Maj} and
references therein). Quantum and braided groups stem from the same
quantum Yang-Baxter equation, which leads to the fact that every
quantum group equipped with an universal $\cal R$-matrix has a
braided-group analog.

The ${\cal R}$-matrix relates, by a similarity transformation, the
coproduct $\Delta_t$ to its opposite $\Delta_t^{op}=\sigma\circ
\Delta_t$, where $\sigma$ is the usual permutation operator of
factors in the tensor product: \be\label{R-matrix} {\cal
R}\Delta_t=\Delta_t^{op}{\cal R}, \ \ \ {\cal R}=\sum {\cal
R}_1\otimes {\cal R}_2\in {\cal H}\otimes{\cal H}. \ee
The Hopf algebra $\cal H$ in which $\Delta_t$ and $\Delta_t^{op}$
are related by such an invertible $\cal R$-matrix is called a
quasi-triangular Hopf algebra. In the case of the twisted \P
algebra, the situation is even simpler, since the $\cal R$-matrix
can be easily shown to be (using (\ref{twist_coproduct}) and
(\ref{R-matrix})) presented as:
\be {\cal R}={\cal F}_{21}{\cal
F}^{-1}=e^{-i\theta^{\mu\nu}P_\mu\otimes P_\nu}\,,\ee
which satisfies ${\cal R}_{12}{\cal R}_{21}=1$. As a result, the
twisted \P algebra is a so-called triangular Hopf algebra.

Turning back to braided groups, the braided permutation (or simply
{\it braiding}) $\Psi$ differs essentially from the bosonic or
graded permutation by the fact that $\Psi\neq\Psi^{-1}$. By finding
the braiding $\Psi$, one can see how quantum groups with universal
$\cal R$-matrix generate nontrivial statistics \cite{ChPr,Maj,ChD}.

Consider now $V$ and $W$, two (co)representation spaces of the
quasi-triangular Hopf algebra $\cal H$. Then the braiding is given
by
\be\label{braiding}\Psi_{V,W}(v\otimes w)=P({\cal
R}\triangleright(v\otimes w))\,,\ee
where $\triangleright$ is the action of ${\cal R}\in {\cal H}\otimes
{\cal H}$, with its first factor acting on $V$ and the second factor
acting on $W$, followed by $P$ - the usual vector-space permutation.
This particular form (\ref{braiding}) for the braiding is obtained
by requiring consistency with the quantum group action, in other
words $\Psi_{V,W}$ has to be intertwiner of representations. Indeed,
if we consider the action of an element of a quasi-triangular Hopf
algebra $h\in {\cal H}$,
\bea h\bullet\Psi(v\otimes w):&=&\Delta (h)\triangleright P({\cal
R}\triangleright(v\otimes w)) =P(\Delta^{op}(h){\cal
R}\triangleright(v\otimes w))\cr &=&P({\cal R}
\Delta(h)\triangleright(v\otimes w)=\Psi(h\bullet(v\otimes
w))\,.\eea

In general for a quasi-triangular Hopf algebra, ${\cal R}_{21}\neq
{\cal R}^{-1}$, consequently $\Psi\neq\Psi^{-1}$ and nontrivial
statistics emerges. However, in the case of a {\it triangular} Hopf
algebra, like the twisted \P algebra, ${\cal R}_{21}= {\cal
R}^{-1}$, leading to $\Psi=\Psi^{-1}$. In other words, $\Psi$ {\it
is symmetric, and not braided}.  Consequently, NC QFTs with twisted
\P symmetry do not experience nontrivial statistics, although the
notion of permutation is deformed using the $\cal R$-matrix.

This rigorous argument based on the analogy between quantum and
braided groups can be made more transparent physically by taking a
concrete example. Consider a free quantum scalar field of mass $m$,
expanded as:
\be
 \phi(x)=\int d\mu({\bf p})\left[a({\bf
p})e^{-ipx}+a^\dagger({\bf p})e^{ipx}\right],\label{free field}\ee
where $d\mu({\bf p})=\frac{d {\bf p}}{(2\pi)^{3/2}2E_p}$ and
$p_0=E_p=\sqrt{{\bf p}^2+m^2}$.

When taking products of quantum fields, one has to deform them by
the use of the inverse of the twist element (\ref{abelian twist}).
At this stage, one is faced with the choice of taking a particular
realization of the momentum generator $P_\mu$. One possibility is to
take the Minkovski space realization, $P_\mu=i\partial_\mu$, in
which case we shall obtain a $\star$-product between exponentials
$$e^{ikx}\star e^{ipx}=e^{ikx} e^{ipx}e^{-\frac{i}{2}k\theta p}$$
(with the notation $k\theta p=k_\mu\theta^{\mu\nu}p_\nu$) and {\it
the usual product between the creation and annihilation operators}.
Another option, since the creation and annihilation operators are
themselves representations of the momentum generator, $${\cal P}_\mu
a({\bf k})=[P_\mu,a({\bf k})]=-k_\mu a({\bf k}),\ \ {\cal P}_\mu
a^\dagger= [P_\mu,a^\dagger({\bf k})]=k_\mu a^\dagger({\bf k})\,,$$
is to take the realization as quantum momentum generator $P_\mu=\int
d^3 k\, k_\mu\, a^\dagger({\bf k})a({\bf k})$, in which case there
will be a $\star$-product between the creation and annihilation
operators, e.g. $$ a^\dagger({\bf k})\star a^\dagger({\bf p})=m\circ
\left(e^{-\frac {i}{2}\theta^{\mu\nu}{\cal P}_{\mu}\otimes {\cal
P}_{\nu}}\right)\left(a^\dagger({\bf k})\otimes a^{\dagger}({\bf
p})\right)=a^\dagger({\bf k})a^\dagger({\bf p})e^{-\frac{i}{2}k
\theta p}\,,$$
but then with {\it the usual multiplication of the exponentials}
\cite{Kulish}.\footnote{There exists still the option of deforming
the multiplication between exponentials and creation and
annihilation operators, as elements of different representation
spaces of the momentum generator. This possibility was explored in
Ref. \cite{AT} with the same results as the other two options
presented here.}

In the case of the Minkovski space realization, the commutation
relations of the creation and annihilation operators are taken, as
usual, as in commutative QFT and the multi-particle states are
defined naturally as
$$|n\rangle=a^\dagger({\bf p}_1)a^\dagger({\bf p}_2)\cdots
a^\dagger({\bf p}_n)|0\rangle,$$ since the multiplication of the
operators is not deformed. As a result, the spin-statistics relation
is obviously preserved.

In the case of the quantum realization of the momentum generator,
the commutation relations of creation and annihilation operators
have to be written as $\star$-commutation relations:
\bea a^\dagger({\bf k})\star a^\dagger({\bf p})=a^\dagger({\bf
p})\star a^\dagger({\bf k})\ e^{-ik \theta p}\label{comm rel cr}\\
a({\bf k})\star a^{\dagger}({\bf p})-e^{ik\theta p}a^{\dagger}({\bf
p})\star a({\bf k})=\delta (\bf{k-p}), \eea
but now the multi-particle states are naturally defined also with
$\star$-product,
\be\label{multipart}|n\rangle_\star=a^\dagger({\bf p}_1)\star
a^\dagger({\bf p}_2)\star\cdots \star a^\dagger({\bf p}_n)|0\rangle.
\ee
According to the new concept of permutation which has to be used in
this case, given by the braiding (\ref{braiding}), a two-particle
state, for example, is symmetric if
\be\label{braid stat} m\circ {\cal F}^{-1}\left(a^\dagger({\bf
k})\otimes a^\dagger({\bf p})\right)=m\circ {\cal
F}^{-1}\Psi\left(a^\dagger({\bf k})\otimes a^\dagger({\bf
p})\right)\,.\ee
This equality can be easily derived by using (\ref{braiding}) and
the obvious relation ${\cal F}_{ 12}={\cal F}_{ 21}^{-1}$:
\bea m&\circ& {\cal F}^{-1}\Psi\left(a^\dagger({\bf k})\otimes
a^\dagger({\bf p})\right)=m\circ {\cal F}^{-1}P\left({\cal
R}\triangleright (a^\dagger({\bf k})\otimes a^\dagger({\bf
p}))\right)\cr &=&m\circ {\cal F}^{-1}P\left( {\cal F}_{21}{\cal
F}^{-1}(a^\dagger({\bf k})\otimes a^\dagger({\bf p})\right) =m\circ
{\cal F}_{ 12}\left(a^\dagger({\bf p})\otimes a^\dagger({\bf
k})\right)\cr &=&m\circ {\cal F}^{-1}_{21}\left(a^\dagger({\bf
p})\otimes a^\dagger({\bf k})\right)=m\circ {\cal
F}^{-1}\left(a^\dagger({\bf k})\otimes a^\dagger({\bf
p})\right)\,.\eea
One can show also that the equality (\ref{braid stat}) is equivalent
to the equality (\ref{comm rel cr}): the equality of the left-hand
sides of the two expression is obvious (by definition), while for
the right-hand sides we have:
\bea m&\circ& {\cal F}^{-1}\Psi\left(a^\dagger({\bf k})\otimes
a^\dagger({\bf p})\right)=m\circ {\cal F}^{-1}P\left({\cal
R}\triangleright (a^\dagger({\bf k})\otimes a^\dagger({\bf
p}))\right)\cr&=&m\circ {\cal
F}^{-1}P\left(e^{-i\theta^{\mu\nu}P_\mu\otimes P_\nu}\triangleright
(a^\dagger({\bf k})\otimes a^\dagger({\bf p}))\right)=m\circ {\cal
F}^{-1}e^{-ik \theta p}\left(a^\dagger({\bf p})\otimes
a^\dagger({\bf k}))\right)\cr &=&a^\dagger({\bf p})\star
a^\dagger({\bf k})\ e^{-ik \theta p}\eea
The equivalence between (\ref{braid stat}) and (\ref{comm rel cr})
means that the $\star$-commutation relation of the creation
operators (\ref{comm rel cr}) ensures the symmetry of the
multi-particle states in accordance with the generalized (deformed)
permutation. At a basic quantum mechanical level, this can be
understood as follows \cite{Fiore}: the deformed commutation
relations of creation and annihilation operators introduce a phase
factor in the new states $|n\rangle_\star$, compared to the usual
ones $|n\rangle$. Thus, the system is defined in the same space of
physical states, by choosing a different representation (by a phase
shift) for the wave functions. Therefore, one has the same
representation of the permutation group and no deformation of
statistics (see also Ref. \cite{yee} for a direct check of this
statement).

We can conclude that NC QFT with twisted \P symmetry does preserve
the spin-statistics relation just as in the commutative
case.\footnote{Starting from the same hypotheses, in Ref.
\cite{bala1} the opposite conclusion was reached.} It is interesting
that the quantum group result is even stronger than the one obtained
in the Lagrangian formulation, which left open the possibility that
NC QFT with light-like noncommutativity may violate the
spin-statistics relation \cite{CNT}. According to the
quantum/braided group approach, any NC QFT with twisted \P symmetry
defined in Ref. \cite{CKT} has the correct spin-statistics relation,
known from usual relativistic QFT.

\section{Quantization of fields and causality condition}

Using the quantization mentioned above and the perturbative
approach, one obtains the causality condition as vanishing of the
commutator of Heisenberg fields out of the light-wedge \cite{Chu}
\be\label{light-wedge}\langle
P_\mu|\left[\phi_H(x),\phi_H(y)\right]_\star|P_\mu\rangle=0\ ,\ \ \
\mbox{for}\ \ (x_0-y_0)^2-(x_3-y_3)^2<0\,, \ee
where it was assumed for simplicity that only
$\theta_{12}=-\theta_{21}\neq 0$, all the other elements of the
$\theta$-matrix being zero. Remark that one cannot obtain a
meaningful commutation relation in operatorial form, but only in
terms of diagonal (physical) matrix elements. The $\star$-products
between fields taken at different points is justified by the fact
that the fields belong to the tensor product of two copies of the
algebra ${\cal A}_\theta$, and after twisting the universal
enveloping algebra of the \P algebra, the elements of different
copies of ${\cal A}_\theta\otimes{\cal A}_\theta$ do not commute,
$$(\phi_1\otimes 1)(1\otimes \phi_2)=\phi_1\otimes \phi_2,\
\mbox{but}\ (1\otimes \phi_2)(\phi_1\otimes 1)=({\cal
R}_2\phi_1)\otimes ({\cal R}_1\phi_2),\ \phi_1,\phi_2\in {\cal
A}_\theta\,.  $$
This noncommutativity can be achieved if we consider a
$\star$-product between fields at different space-time points,
$\phi(x)\star\phi(y)$ \cite{oeckl,Kulish} (a similar result was
arrived at in \cite{q-shift}, using the concept of quantum shift).
Consequently, the Wightman functions, for example, will be
rigorously written with $\star$-products, $\langle0| \varphi(x_1)
\star \varphi(x_2) \star\cdots \star \varphi(x_n)|0\rangle$.

The light-wedge causality condition is a mark of the infinite
nonlocality in the noncommutative directions, leading, among other
things, to the UV/IR mixing. This nonlocality is supported by dual
twisted \P group, generated by the parameters of global
transformations, $\Lambda^\mu_\nu, a^\mu$. It was shown that, while
the parameters of rotations still commute, the parameters of finite
translations do not commute \cite{oeckl,Gonera}:
\be\label{finite
translations}[a^\mu,a^\nu]=i\theta^{\mu\nu}-i\Lambda^\mu_\alpha\Lambda^\nu_\beta\theta^{\alpha\beta}\,,\
\ \  [\Lambda^\mu_\nu, a^\alpha]=[\Lambda^\mu_\alpha,\Lambda^\nu
_\beta]=0\,.\ee
Moreover, in Ref. \cite{Gonera} it was proved that the global
twisted \P group is physically equivalent to the stability group of
the $\theta$-matrix, i.e. $O(1,1)\times SO(2)$, which justifies the
axiomatic approach based on this group and the derived consequences
(the extensions of the spin-statistics and CPT theorems)
\cite{LAG_ax,axiomatic}.

The quantization of NC quantum fields in a theory with twisted \P
invariance is heavily based on the commutative example, in the sense
that the mode-expansion of the free fields is taken exactly from the
commutative theory as in (\ref{free field}) and then one adjusts the
commutation relations of the creation and annihilation operators, or
the commutation relations of exponentials, or the commutation
relations of creation/annihilation operators and exponentials, to
make the product of quantum fields {\it twisted} \P covariant. An
alternative approach is to start from a canonical quantization
prescription. Once settled the fact that the $\star$-product between
two fields at different space-time points is necessary, one has to
impose a canonical quantization condition, in terms of
$\star$-commutators of fields and their canonically conjugated
momenta. According to the present-day wisdom, one would impose a
condition which takes into account the nonlocality of the theory in
the NC directions, i.e.
\be\label{canonical
NC}\left[\phi(x),\pi(y)\right]_\star=\delta(x_3-y_3)\,,\ee
for the choice of the $\theta$-matrix adopted in the beginning of
this section. In that case, the corresponding locality condition is
\be\left[\phi(x),\phi(y)\right]_\star=0 ,\ \ \ \mbox{for}\ \
(x_0-y_0)^2-(x_3-y_3)^2<0\,. \ee
Remark that, according to the idea of the twist, the individual free
fields are representations of the usual \P algebra. The insertion of
the $\star$-product in constructing the observables makes them then
rotationally non-invariant, due to the presence of \th, but still
twisted \P invariant, while the r.h.s. is symmetric under the
stability group of $\theta$ and this is physically consistent, as
shown in Ref. \cite{Gonera}.

An alternative (see, e.g., Refs. \cite{Abe,FW}) would be to take in
the r.h.s. of (\ref{canonical NC}) a three-dimensional
$\delta$-function,
\be\label{canonical
usual}\left[\phi(x),\pi(y)\right]_\star=\delta(\bf {x-y})\,,\ee
or equivalently a locality condition expressed in terms of
light-cone, not light-wedge:
\be\left[\phi(x),\phi(y)\right]_\star=0 ,\ \ \ \mbox{for}\ \
(x_0-y_0)^2-(\bf{x-y})^2<0\,. \ee
If in the r.h.s we choose an expression which is rotationally
invariant (like $\delta(\bf {x-y})$) or Poincar\'e-invariant (like
the light-cone), this automatically {\it forces} the l.h.s. into
having the same higher symmetry. As a result, a mechanism will
appear to {\it cancel} the $\star$-product (e.g., deformed
commutation relation of creation and annihilation operators, taken
at the same time with $\star$-products of exponentials) and the
emerging theory will have usual, and not twisted, Poincar\'e
invariance.

We consider that the key of the issue is the noncommutativity of the
finite translations (\ref{finite translations}) when Lorentz
transformations are also involved.

\section{Twisted gauge theories}

It is natural to ask whether the concept of {\it twist} provides a
{\it symmetry principle} in constructing NC field theories, i.e. any
symmetry that NC field theories may enjoy, be it space-time or
internal symmetry, global or local, should be formulated as a
twisted symmetry. To investigate this issue, let us consider the
possibility of formulating twisted gauge theories.

NC gauge theories have been traditionally formulated by using the
Weyl-Moyal correspondence \cite{Hayakawa}, i.e. replacing in the
action for a commutative gauge theory all the products by
$\star$-products. It turns out that the gauge symmetry that these
theories enjoy is strongly constrained \cite{nogo} due to the
presence of the $\star$-products: only $U_\star(n)$-gauge groups
close (not, e.g., $SU_\star(n)$) and a no-go theorem applies,
stating that i) matter fields can only be in (anti)-fundamental,
adjoint or singlet states and ii) for any NC group consisting of
several group factors, matter fields can transform nontrivially
under at most two groups.

Any reasonable model building is impossible without circumventing
somehow this no-go theorem. Therefore, a twisted symmetry principle
which would lead to NC gauge theories with any local internal group
of symmetry and admitting any representations would represent a
remarkable progress. The idea to twist the coproduct of the gauge
generators with the same Abelian twist (\ref{abelian twist}) was
explored in Refs. \cite
{Vassilevich}. Concretely, the hope was that,
if an individual field transforms under infinitesimal gauge
transformations as
\be\delta_\alpha \phi(x)=\alpha(x)\phi(x),\ \
\alpha(x)=i\alpha^a(x)T_a,\ \ [T_a,T_b]=if_{abc}T_c\,,\ee
then the $\star$-product of two fields will transform as
\cite{Vassilevich}:
\be\label{product gauge transf} \delta_\alpha (\phi_1(x)\star
\phi_2(x))=i\alpha^a(x)[(\phi_1
T^{(1)}_a)\star\phi_2+\phi_1\star(T^{(2)}_a\phi_2)]\,.\ee

However, this could be true only in the case in which once a field
transforms in a representation of the gauge group, its derivatives
of any order would transform in the same representation, i.e.
\be\label{deriv transf}
\delta_\alpha((-i)^nP_{\mu_{1}}...P_{\mu_{n}}\phi(x))=\delta_\alpha(\partial_{\mu_{1}}...\partial_{\mu_{n}}\phi(x))
=\alpha(x)(\partial_{\mu_{1}}...\partial_{\mu_{n}} \phi(x)), \ee
because
\bea&&\delta_\alpha (\phi_1\star \phi_2)
=m_\star\circ\Delta_t(\delta_\alpha(x))(\phi_1\otimes \phi_2) =
m\circ\Delta_0(\delta_\alpha){\cal F}^{-1}(\phi_1(x)\otimes
\phi_2(x))\cr
&&= m\circ(\delta_\alpha\otimes 1 +1\otimes
\delta_\alpha)\times\left[\phi_1\otimes \phi_2+ \frac{i}{2}
\theta^{\mu\nu}\left(\partial_\mu\phi_1\otimes
\partial_\nu\phi_2\right)+\cdots\right].\eea
Clearly, (\ref{deriv transf}) is against the very concept of gauge
principle, and thus this approach in constructing twisted gauge
theories is not consistent \cite{CT}.

However, the idea should not be abandoned, since one could replace
the usual derivatives by covariant derivatives, in which case indeed
we have
$$\delta_\alpha(D_{\mu_{1}}...D_{\mu_{n}}\phi)=\alpha(x)(D_{\mu_{1}}...D_{\mu_{n}}\phi).$$
This suggests us to introduce a non-Abelian twist element \cite{CTZ}
\be\label{nonabelian twist} {\cal T} = {\rm Exp}\left(-\frac{i}{2}
\theta^{\mu\nu}D_\mu\otimes D_\nu+{\cal O}(\theta^2)\right), \ee
where Exp is a power series expansion which reduces to the usual
Abelian twist in the absence of gauge fields. The non-Abelian twist
element has to satisfy a twist condition like (\ref{twist_eq}),
which would ensure the associativity of the corresponding
$\star$-product. So far, in spite of having considered the most
general ansatz, a power series of the form (\ref{nonabelian twist}),
satisfying the twist condition, has not been found \cite{CTZ}. It is
intriguing that the {\it external} \P symmetry and the {\it
internal} gauge symmetry cannot be unified under a common twist. The
situation is somehow reminiscent of the Coleman-Mandula theorem and
it is interesting to speculate whether supersymmetry may reverse the
situation.

\section{Conclusions}

We emphasize once more that the NC QFT formulated with the
Weyl-Moyal correspondence has twisted Poincar\'e symmetry, and this
symmetry is in full accord with the preservation of the
spin-statistics relation and the infinite nonlocality in the
noncommutative directions (leading to phenomena like the UV/IR
mixing), previously shown and studied in the literature. As a
result, commutative and noncommutative QFT could never be identical.
The hope is to find a twisted symmetry principle for NC QFT, which
would lead to consistent formulations of NC gauge theories and NC
gravity.

\section*{Acknowledgements}
We would like to thank Masud Chaichian, Petr Kulish, Kazuhiko
Nishijima and Peter Pre\v{s}najder for discussions, on different
occasions, about the topics elaborated on in this talk.

\end{document}